\documentclass[prb,preprint,nofootinbib,superscriptaddress]{revtex4-2}
\usepackage{amssymb,mathrsfs,xcolor}
\usepackage{amsmath,amsfonts,graphicx}
\usepackage{bm} 

\newcommand{\nn}{\nonumber\\}

\newcommand{\la}{\langle}
\newcommand{\ra}{\rangle}
\newcommand{\ben}{\begin{displaymath}}
\newcommand{\een}{\end{displaymath}}
\newcommand{\be}{\begin{equation}}
\newcommand{\ee}{\end{equation}}
\newcommand{\bea}{\begin{eqnarray}}
\newcommand{\eea}{\end{eqnarray}}
\newcommand{\eqn}[1]{\label{#1}}
\newcommand{\eq}[1]{Eq.~(\ref{#1})}

\begin{document}
\title{New derivation of Time-Independent Perturbation Theory}
\author{A.\ N.\ Kvinikhidze}
\email{sasha\_kvinikhidze@hotmail.com}
\affiliation{Andrea Razmadze Mathematical Institute of Tbilisi State University, 6, Tamarashvili Str., 0186 Tbilisi, Georgia}
\affiliation{
College of Science and Engineering,
 Flinders University, Bedford Park, SA 5042, Australia}
\author{B.\ Blankleider}
\email{boris.blankleider@flinders.edu.au}
\affiliation{
College of Science and Engineering,
 Flinders University, Bedford Park, SA 5042, Australia}

\date{\today}

\begin{abstract}
We propose a new derivation of Time-Independent Perturbation Theory (PT) that has a fundamental advantage over the  usual derivations presented in textbooks on Quantum Mechanics (QM): it is simpler and much shorter. As such, it can provide an easier and quicker way for students to learn PT, than afforded by current methods. In spite of that, our approach does not require the potentials to be energy independent or the inverse free Green function $G_0^{-1}(E)$ to be a linear function of energy $E$, as is the case in QM, and can be applied directly to various extensions of QM including Relativistic QM, the Bethe-Salpeter equation, and all kinds of quasipotential approaches in Quantum Field Theory.
\end{abstract}

\maketitle
\newpage

\section{Introduction}

 In Quantum Mechanics (QM), almost none of the physically interesting problems can be solved exactly. This makes Time-Independent Perturbation Theory (PT) a fundamental part of QM, as it allows one to approximate, to any level of accuracy, the energies and wave functions of any potential $K$ that differs not too greatly from a potential $K_0$ whose energies and wave function are known. For this reason, PT features as a major topic in all textbooks on QM. {What is noteworthy, is that the approach used to derive the expressions of PT is standard: one starts with the Schr\"{o}dinger equation for potential $K$ expressed as $K_0$ plus a ``perturbation" $K_1$ (thus $K=K_0+K_1$), one then introduces a small parameter $\lambda$ into $K_1$ (thus $K=K_0+\lambda K_1$), and then one expands all $\lambda$-dependent terms into a power series in $\lambda$. By utilising the orthonormality and completeness of eigenstates, the coefficients of the power series 
can then be determined by various means.  
 At its core, this approach relies on the potentials $K_0$ and $K_1$ being energy independent, thereby resulting in a complete set of orthonormal eigenstates that are essential for the derivation.}
 
In this paper we would like to introduce a different way of deriving PT that has a significant advantage over the standard one just described: it is simpler and much shorter. Moreover, our formulation remains valid for extensions of QM, like Relativistic QM {(see \cite{Rizov:1984uk} and reference therein)}, the Bethe-Salpeter (BS) equation \cite{Salpeter:1951sz, Gell-Mann:1951ooy}, and all kinds of quasipotential approaches in Quantum Field Theory \cite{Logunov:1963yc,Blankenbecler:1965gx,Gross:1969rv,*Gross:1982nz,*Gross:1982ny}, where the potentials can be energy dependent and where the inverse free Green function $G_0^{-1}(E)$ is not restricted to be the operator $E-H_0$ as in QM, but can be any function of energy $E$.

\section{New derivation}

\subsection{Preliminaries}

{Textbooks on QM derive PT in a variety of ways, however they all share the same overall approach (see for example \cite{schiff:quantum, Shankar1994-aj,Merzbacher1997-le,Gasiorowicz2003-bj,Griffiths2018-pn,Sakurai2020-nt,Zettili2022-eo}). For completeness, we provide a quick summary of this standard approach in Appendix A.} Here we describe an alternative approach. The method is much simpler than the one we previously proposed \cite{Kvinikhidze:2003ft}.

We consider a system subject to a potential $K=K_0+K_1$ where $K_1$ is considered to be a perturbation to a potential $K_0$ for which the bound state problem has been completely solved.  {Otherwise we make no assumptions about the energy dependence of $K_0$, $K_1$, or of the Green function $G_0(E)$ describing the free propagation of particles in this system. } In particular, we would like to find the solution of the bound state equation for the perturbed system,
\be
[G^{-1}_0(E_n)-K_0 -K_1] |\psi_n \ra = 0 ,  \eqn{GSE}
\ee
when the unperturbed system has a bound state described by the bound state equation
\be
[G^{-1}_0({\cal E}_n)-K_0 ] |\phi_n \ra = 0    \eqn{GSEu}
\ee
where not only ${\cal E}_n$ and $|\phi_n\ra$ are known, but so is the unperturbed Green function  $G_u$ defined as the solution of the inhomogeneous equation,
\be
G_u  = G_0+G_0 K_0 G_u . \eqn{Gu}
\ee
In the case of QM, this statement of the PT problem is equivalent to the one given in the standard approach described in Appendix A. Indeed, \eq{GSE} is just a rearranged version of \eq{SE},  and \eq{GSEu} is just a rearranged version of \eq{SEu}, and knowing {\it all}  the solutions of the unperturbed Schr\"{o}dinger equation, \eq{SEu}, as is assumed in the standard approach, implies that $G_u$ is known, since in that case,
\be
G_u(E) = \sum_m \frac{| \phi_m\ra\la \phi_m|}{E - {\cal E}_m},    \eqn{GuQM}
\ee
which follows from the definition of $G_u$ and the completeness relation \eq{comp}. However, in the general case applicable to various extensions of QM, our statement of the PT problem is less restrictive, it allows $K_0$ and $K_1$ to be energy dependent and is applicable to any functional form of  $G^{-1}_0(E)$. Of course in the general case (for example in the BS approach), \eq{GuQM} does not apply and $G_u(E)$ would need to be found by solving \eq{Gu}.

{Being defined by \eq{Gu},  $G_u$ is an interacting Green functions driven by potential $K_0$.  A general feature of interacting Green functions is that they exhibit poles at energies corresponding to  bound states or resonances produced by their driving potentials, with the residue consisting of an outer product of their corresponding bound state wave functions.} Thus, if the unperturbed potential $K_0$ gives rise to a bound state at energy ${\cal E}_n$, one can express $G_u(E)$ as
\be
G_u(E) = \frac{|\phi_n\ra\la\phi_n|}{E-{\cal E}_n} + G_u^b(E)    \eqn{Gpole}
\ee
where $|\phi_n\ra$ is the corresponding  bound state wave function and $G_u^b(E)$ is a ``background" term.
{Note that \eq{Gpole} implies the normalisation condition $\la\phi_n|\partial(G^{-1}_0-K_0)/\partial E|\phi_n\ra=1$ which in the case of QM, reduces to $\la\phi_n|\phi_n\ra = 1$.}
In writing \eq{Gpole} we have assumed the simplest case of no degeneracy for energy ${\cal E}_n$, but our presentation is easily extended to the case of degeneracy simply by replacing the outer product $|\phi_n\ra\la\phi_n|$ with ths sum $\sum_j|\phi^j_n\ra\la\phi^j_n|$ where $|\phi^j_n\ra$ is the $j$'th degenerate state. In QM, \eq{Gpole} follows from \eq{GuQM} and is a consequence of completeness, while in Quantum Field Theory  it is usually related the vacuum expectation of the time-ordered product of fields, $G_u(E) = \la 0| T\psi\psi\bar\psi\bar\psi|0\ra$.

\subsection{Derivation}

Using \eq{Gpole} as representative of the general case where the potentials can be energy dependent,  the derivation of the perturbed bound state energy $E_n$ and corresponding bound state  function $|\psi_n\ra$, is remarkably short. {The key idea is to 
use  \eq{Gu} and \eq{Gpole} to write \eq{GSE} in the form\footnote{$G^b\equiv [1-G^b_u(E_n) K_1(E_n)]^{-1}$ should be understood as the solution of the equation $G^b=G^b_u+G^b_u K_1G^b$ or as the infinite series of \eq{K1n}.}
\be
|\psi_n\ra =\frac{[1-G^b_u(E_n) K_1(E_n)]^{-1}|\phi_n\ra \la \phi_n |}{E_n-{\cal E}_n}  K_1(E_n)|\psi_n\ra ,\eqn{pap-wf}
\ee
which is a bound state equation with a separable kernel, and which can therefore be solved algebraically. Indeed, the shortness of the derivation is due directly to the trick of reducing the PT problem to that of an equation with a separable kernel, a trick that does not appear to have been noticed previously.
In particular, \eq{pap-wf} can be used to write the equation for the scalar combination $\la \phi_n| K_1|\psi_n\ra$,}
\be
\la \phi_n| K_1|\psi_n\ra=\frac{\la\phi_n| K_1(1-G^b_u K_1)^{-1}|\phi_n\ra }{E_n-{\cal E}_n} \la\phi_n| K_1|\psi_n\ra.
\ee
This is an algebraic equation for $\la\phi_n| K_1|\psi_n\ra $ which has a nontrivial solution only if
\be
E_n={\cal E}_n+ \la\phi_n| K_1(E_n)[1-G^b_u(E_n) K_1(E_n)]^{-1}|\phi_n\ra.    \eqn{En}
\ee
{To solve \eq{pap-wf} for $|\psi_n\ra$ unambiguously, a normalisation condition is needed, which we choose to correspond to the solution
\be
|\psi_n\ra = [1-G^b_u(E_n) K_1(E_n)]^{-1}|\phi_n\ra , \eqn{psin}
\ee
as this choice simplifies the PT expressions. {There is no problem to renormalise $|\psi_n\ra$  in order to construct amplitudes for scattering processes where the bound states are involved \cite{Kvinikhidze:2003ft}. Note also that in QM,  \eq{psin} implies $\la \phi_n|\psi_n\ra = 1$, in agreement with the normalisation chosen in standard derivations of PT \cite{Griffiths2018-pn,Sakurai2020-nt}.}

The simplicity and the shortness of the derivation of \eq{En} and \eq{psin} constitute the main results of this paper. They provide exact expressions for $E_n$ and $|\psi_n\ra$, albeit as functions of the yet undetermined exact energy $E_n$. As such, it is possible that they could be solved numerically by iteration, starting with ${\cal E}_n$ as the first estimate of $E_n$. However, they can also be used to generate $E_n$ and $|\psi_n\ra$ at each order of $K_1$ in perturbation theory, by  introducing the expansion parameter $\lambda$ via the replacement $K_1\rightarrow \lambda K_1$ and expanding \eq{En} and \eq{psin} in powers of $\lambda$. To pursue this procedure it may be convenient to expand \eq{En} and \eq{psin} in powers of $K_1$ beforehand, using
\be
[1-G^b_u(E_n) K_1(E_n)]^{-1}= \sum_{s=0} [G^b_u(E_n) K_1(E_n)]^s .  \eqn{K1n}
\ee

\section{Discussion}

We have presented a new derivation of PT that applies not only to QM, but also to various extensions of QM for which the unperturbed potential $K_0$ and the perturbation potential $K_1$ can be energy dependent, and for which the inverse free Green function $G_0^{-1}(E)$ can be a nonlinear function of $E$. However, the main feature of the new approach is its simplicity and the shortness of its derivation, resulting in the expressions \eq{En} and \eq{psin} for the perturbed energy and wavefunction $E_n$ and $|\psi_n\ra$, respectively.

{Although  \eq{En} and \eq{psin}  were derived for the case where the only difference between the perturbed and unperturbed systems is the presence or absence of the perturbation potential $K_1$, our approach enables the use of PT for a yet more general case, where in addition to $K_1$, the perturbed and unperturbed systems differ by having different free Green functions. Denoting by $G_1(E)$ and $G_0(E)$ the free Green functions of the perturbed and unperturbed systems, respectively, it is clear that in the derivation of Sec.\ II, all that needs to be done to describe this more general case, is to replace \eq{GSE} with
\begin{align}
[G^{-1}_1(E_n)-K_0 -K_1] |\psi_n \ra = 0,
 \eqn{GSE1}
\end{align}
which can of course be written as
\begin{align}
[G^{-1}_0(E_n)-K_0 - \tilde K_1] |\psi_n \ra = 0,
 \eqn{GSE2}
\end{align}
where $\tilde K_1 = K_1+\Delta$ and
\be
\Delta = G^{-1}_0(E_n) - G^{-1}_1(E_n)
\ee
is an energy-dependent addition to the perturbation potential $K_1$. Thus, to obtain a PT for this more general case, all one needs to do is to use $\tilde K_1$ instead of $K_1$ in \eq{En} and \eq{psin}. Such a PT would be useful, for example, to describe a relativistic perturbation where the inverse free Green function $G^{-1}_1(E)$ of the perturbed system is a non-linear function of $E$, while the completely solved unperturbed system uses the standard linear inverse free Green function $G^{-1}_0(E)= E-H_0$ of nonrelativistic QM. 
}

{It is noteworthy that in the case where $K_0$ is energy independent, so that the unperturbed Green function $G_u$ is given by \eq{GuQM}, but where $K_1(E)$ is still energy dependent, our perturbed wave function can be written in the form given by Sakurai and Napolitano for QM (see Eq.\ (5.34) of Ref.\ \cite{Sakurai2020-nt}),
\be
|\psi_n\ra = [1-G^b_u({\cal E}_n) \left(K_1(E_n)-E_n+{\cal E}_n\right)]^{-1}|\phi_n\ra . \eqn{sakurai 5.34}
\ee
The distinctive feature of this expression is that the argument of $G^b_u({\cal E}_n)$ is the unperturbed energy ${\cal E}_n$, not $E_n$. Equation (\ref{sakurai 5.34}) is derived in Appendix B. In this case the perturbed energy, \eq{En}, can similarly be written as
\be
E_n={\cal E}_n  + \la\phi|  K_1(E_n) \left[1- G^b_u({\cal E}_n)\left(K_1(E_n)-E_n+{\cal E}_n\right)\right]^{-1}|\phi\ra     .
\eqn{Sak-E}
\ee}

To conclude, we note that the efficiency of our approach can be demonstrated explicitly for the case of QM for which 
\be
G^b_u(E) = \sum_{m\ne n} \frac{| \phi_m\ra\la \phi_m|}{E - {\cal E}_m}.    \eqn{GubQM}
\ee
Substituting this expression into \eq{K1n}, the first few term of \eq{En} and \eq{psin} become 
\begin{subequations}
\begin{align}
E_n &= {\cal E}_n +   \la\phi_n| K_1|\phi_n\ra + 
\sum_{m\ne n} \frac{|\la\phi_m |K_1|\phi_n\ra|^2}{E_n-{\cal E}_m}  +\dots \\[2mm]
|\psi_n\ra &=  |\phi_n\ra +  \sum_{m\ne n}  \frac{ \la \phi_m|K_1 |\phi_n\ra}{E_n-{\cal E}_m} |\phi_m\ra \nn[2mm]
&+ \sum_{k\ne n} \sum_{m\ne n}  \frac{\la \phi_m| K_1|\phi_k\ra\la \phi_k| K_1|\phi_n\ra}
{(E_n - {\cal E}_m)(E_n - {\cal E}_k)} |\phi_m\ra
 + \dots
\end{align}
\end{subequations}
which immediately gives the perturbation series \eq{sakurai} of the standard approach.
{This constitutes a much easier and shorter derivation of PT, and its main results, than that of the standard approach used in all current  textbooks on QM. As such, our derivation may present students of QM a simpler and quicker way to learn PT.}

\acknowledgments

 A.N.K. was supported by the Shota Rustaveli National Science Foundation (Grant No. FR17-354).

\appendix
\section{Standard approach to PT in QM}

In QM, the standard way of framing the task of PT is as follows (see for example Griffiths and Schroeter \cite{Griffiths2018-pn}). Suppose one knows {\it all} the solutions of the Schr\"{o}dinger equation
\be
H_u |\phi_n\ra = {\cal E}_n |\phi_n\ra       \eqn{SEu}
\ee
where $H_u$ is some ``unperturbed" Hamiltonian
\be
H_u = H_0 + K_0,
\ee
$H_0$ being the kinetic energy operator and $K_0$ being the potential. In almost all presentations it is assumed that the energy levels are discrete, so that $n$ is an integer and the solutions $|\phi_n\ra$ satisfy the orthonormality and completeness conditions 
\begin{subequations}
\begin{align}
\la \phi_n | \phi_m\ra &= \delta_{nm},\\
\sum_m |\phi_m\ra\la \phi_m| &= I. \eqn{comp}
\end{align}
\end{subequations}
For simplicity, we shall also make this assumption while recognising that in many practical applications one also needs to include scattering states \cite{doi:10.1119/10.0001428}.
Now suppose a perturbation is introduced into the system by adding a potential $K_1$. The problem then is to determine the solutions of 
\be
H |\psi_n\ra = E_n |\psi_n\ra    \eqn{SE}
\ee
where $H$ is the Hamiltonian with the perturbation included:
\be
H = H_u + K_1.
\ee
The standard way to solve this problem is to introduce a parameter $\lambda$  such that
\be
H(\lambda) = H_u +\lambda K_1.
\ee
One then solves the Schr\"{o}dinger equation
\be
(H_u + \lambda K_1) |\psi_n(\lambda)\ra = E_n(\lambda) |\psi_n(\lambda)\ra
\ee
by making a power expansion in $\lambda$,
\begin{subequations}
\begin{align}
| \psi_n(\lambda) \ra &=  |\phi_n\ra +\lambda |\psi^1_n\ra+\lambda^2 |\psi^2_n\ra +\dots ,\\
E_n(\lambda) &= {\cal E}_n + \lambda E_n^1 + \lambda^2 E_n^2 + \dots .   \eqn{Enps}
\end{align}
\end{subequations}
The coefficients of these power series are then determined by various means. After a substantial amount of  algebra, and assuming there is no degeneracy, one obtains the perturbation series \cite{Griffiths2018-pn,Sakurai2020-nt} (shown explicitly up to second order in $K_1$)
\begin{subequations} \eqn{sakurai}
\begin{align}
E_n &= {\cal E}_n +   \la\phi_n| K_1|\phi_n\ra + 
\sum_{m\ne n} \frac{|\la\phi_m |K_1|\phi_n\ra|^2}{{\cal E}_n-{\cal E}_m}  +\dots \\[2mm]
|\psi_n\ra &=  |\phi_n\ra +  \sum_{m\ne n}  \frac{ \la \phi_m|K_1 |\phi_n\ra}{{\cal E}_n-{\cal E}_m} |\phi_m\ra 
- \sum_{m\ne n}  \frac{ \la \phi_m|K_1 |\phi_n\ra\la \phi_n|K_1 |\phi_n\ra}{({\cal E}_n-{\cal E}_m)^2} |\phi_m\ra
\nn[2mm]
&+ \sum_{k\ne n} \sum_{m\ne n}  \frac{\la \phi_m| K_1|\phi_k\ra\la \phi_k| K_1|\phi_n\ra}
{({\cal E}_n - {\cal E}_m)({\cal E}_n - {\cal E}_k)} |\phi_m\ra
 + \dots
\end{align}
\end{subequations}

\section{Derivation of \eq{sakurai 5.34}}

Assuming that $K_0$ (but not $K_1$) is energy independent, the unperturbed Green function $G_u(E)$ is given by \eq{GuQM}  and its background part,  $G^b_u(E)$,  is given by \eq{GubQM}. Defining  
\be
\Delta_n = E_n - {\cal E}_n,
\ee
one can write down the Taylor expansion of $G^b_u(E_n)$ around the point ${\cal E}_n$ as
\be
G^b_u(E_n)=G^b_u({\cal E}_n) - \Delta_n [G^b_u({\cal E}_n)]^2 + \Delta^2_n [G^b_u({\cal E}_n)]^3+\dots 
\ee
Writing $K_1\equiv K_1(E_n)$, \eq{psin} then becomes
\begin{align}
|\psi_n\ra &= [1-G^b_u(E_n) K_1]^{-1}|\phi_n\ra \nn
&= [1+G^b_u( E_n) K_1+G^b_u( E_n) K_1G^b_u( E_n) K_1+\dots]|\phi_n\ra
\nn
& = [1+G^b_u( {\cal E}_n) K_1- \Delta_n [G^b_u({\cal E}_n)]^2 K_1+G^b_u( {\cal E}_n) K_1G^b_u( {\cal E}_n) K_1+\dots]|\phi_n\ra
\nn
& = [1+G^b_u( {\cal E}_n) K_1+G^b_u( {\cal E}_n) (K_1-\Delta_n)G^b_u( {\cal E}_n) K_1+\dots]|\phi_n\ra
\nn
& = [1-G^b_u( {\cal E}_n) (K_1{ -\Delta_n})+G^b_u( {\cal E}_n) (K_1-\Delta_n)G^b_u( {\cal E}_n) (K_1{ -\Delta_n})+\dots]|\phi_n\ra
\nn
& = [1-G^b_u( {\cal E}_n) (K_1 -\Delta_n)]^{-1} |\phi_n\ra,
\end{align}
where we have used the fact that $G^b_u( {\cal E}_n) |\phi_n\ra = 0$.

\bibliography{/Users/phbb/Physics/Papers/refn} 

\end{document}